%
\ifx\mnmacrosloaded\undefined \input mn\fi
%
\newif\ifAMStwofonts

\ifCUPmtplainloaded \else
  \NewTextAlphabet{textbfit} {cmbxti10} {}
  \NewTextAlphabet{textbfss} {cmssbx10} {}
  \NewMathAlphabet{mathbfit} {cmbxti10} {} 
  \NewMathAlphabet{mathbfss} {cmssbx10} {} 
  \ifAMStwofonts
    \NewSymbolFont{upmath} {eurm10}
    \NewSymbolFont{AMSa} {msam10}
    \NewMathSymbol{\upi}     {0}{upmath}{19}
    \NewMathSymbol{\umu}     {0}{upmath}{16}
    \NewMathSymbol{\upartial}{0}{upmath}{40}
    \NewMathSymbol{\leqslant}{3}{AMSa}{36}
    \NewMathSymbol{\geqslant}{3}{AMSa}{3E}

     \let\ge=\geqslant
  \else
    \def\umu{\mu}
    \def\upi{\pi}
    \def\upartial{\partial}
  \fi
\fi


\pageoffset{-2.5pc}{0pc}

\loadboldmathnames



\pagerange{1--11}    
\pubyear{1998}
\volume{000}

\begintopmatter  

\title{On the Polarization of H$\alpha$ Scattered by
Neutral Hydrogen in Active Galactic Nuclei}
\author{Hee-Won Lee and Jong-Hyeun Yun}
\affiliation{Department of Astronomy, Seoul National University,
Seoul, 151-742, Korea}

\shortauthor{Lee and Yun}
\shorttitle{Polarization of H$\alpha$ in Active Galactic Nuclei}


\acceptedline{Accepted 1998 July 28. Received 1998 July 27;
  in original form 1998 July 26}

\abstract {Raman scattering by atomic hydrogen converts the UV continuum 
around Ly$\beta$
into optical continuum around H$\alpha$, and the basic atomic physics has been
discussed in several works on symbiotic stars. We propose that
the same process may operate in active galactic nuclei (AGN) and
calculate the linear polarization of the broad emission lines
Raman-scattered by a high column neutral hydrogen compnent. The conversion
efficiency of the Raman scattering
process is discussed and the expected scattered flux is computed using the
spectral energy distribution of an AGN given by a typical power law. The 
high column H\,{\sc i} component in AGN is suggested by many observations 
encompassing radio through UV and X-ray ranges.

When the neutral hydrogen component with a column density $\sim 10^{22}\ 
{\rm cm^{-2}}$ is present around the
active nucleus, it is found that the scattered H$\alpha$ is characterized by
a very broad width $\sim 20,000\ 
{\rm km~s^{-1}}$ and that the strength of the polarized flux is comparable to
that of the electron-scattered flux expected from a conventional unified model
of narrow line AGN. The width of the scattered flux is mainly determined
by the column density of the neutral scatterers where the total scattering
optical depth becomes of order unity.  The asymmetry
in the Raman scattering cross section around Ly$\beta$ introduces red 
asymmetric polarized profiles around H$\alpha$. The effects of the
blended Ly$\beta$ and O\,{\sc vi} 1034 doublet are also investigated.

We briefly discuss the spectropolarimetric observations performed
on the Seyfert galaxy IRAS~110548-1131 and the narrow line radio
galaxy Cyg~A. Several predictions regarding the scattering by a
high column neutral hydrogen component in AGN are discussed.}


\maketitle  

\section{Introduction}

The spectra of active galactic nuclei (hereafter AGN) are characterized by
many prominent emission lines. The width of the emission lines
is one of the important parameters to classify these active galaxies.
For example, type 1 Seyfert galaxies possess both broad and narrow emission
lines and type 2 Seyfert galaxies exhibit only narrow emission lines
(e.g. Osterbrock 1989). Radio galaxies are also classified into broad line radio
galaxies and narrow line radio galaxies in a similar way. The width of
the broad emission lines (BEL's) often ranges up to
several thousand ${\rm km~s^{-1}}$ and the narrow emission lines (NEL's)
show an order of magnitude smaller width in both type Seyfert galaxies.
 
A unification model of Seyfert galaxies suggests that
Seyfert 2 nuclei also have the broad emission line region (BELR), which
is blocked from the observer's line of sight. In this unification
scheme, the BEL photons can be scattered from other lines of sight
to enter the observer's line of sight. Because the scattered
component is expected to be highly polarized, spectropolarimetry
plays an important role to test the unification models (e.g. Antonucci 1993).
 
A similar behavior is
seen in broad absorption line quasars, in which the broad absorption
trough is partially filled by photons scattered from other lines of
sight. These photons are believed to be scattered resonantly by a fast moving
medium just outside the BELR, and several spectropolarimetric observations
using large telescopes such as the Keck 10 m telescope have been performed
(e.g. Cohen et al. 1995, Goodrich \& Miller 1995, Lee \& Blandford 1997).
 
Spectropolarimetry of Seyfert 2 galaxies performed by several researchers
confirmed that the BEL's are shown in the polarized flux
as expected from the unification scheme (e.g. Antonucci \& Miller 1985,
Miller \& Goodrich 1990, Antonucci 1993). The constant
position angle and degree of polarization favor strongly the model,
in which the electron plasma located in high latitude scatters the radiation
originated from the BELR into the observer's line of sight.
 
Tran (1995 a,b,c) presented the spectropolarimetric data on a number of
Seyfert 2 galaxies, which exhibit broad emission lines in the polarized flux.
He noted that the Balmer lines are more highly polarized than the adjacent
continuum even after all the corrections due to the star light and the
interstellar medium. He proposed an explanation by assuming that there is still
unpolarized continuum having a flat spectra not included in the correction
procedures. This component provides more dilution to
the continuum leaving the lines more polarized. However, higher stellar
contribution as suggested by Koski (1978) would result in wavelength
independent polarization (Antonucci, 1993).
 
Another interesting and important spectropolarimetric observation is
presented by Ogle et al. (1997), who observed the narrow line
radio galaxy Cyg~A. They found that the H$\alpha$ polarized flux
has a very broad width exceeding $20000\ {\rm km~s^{-1}}$.
Young et al. (1993) also found very broad polarized H$\alpha$ from
their spectropolarimetry of the Seyfert 2 galaxy IRAS~110548-1131,
where the full width at zero intensity is $16800\ {\rm km~s^{-1}}$.
 
The X-ray observations of Seyfert galaxies have been performed to
elucidate the physical properties of the highly ionized component
plausibly related with the scattering agents of the Seyfert 2 galaxies.
The warm absorbers having a column density of $N_{H}\sim 10^{21}-10^{22}
\ {\rm cm^{-2}}$ are invoked to explain the soft X-ray observations of Seyfert
galaxies. On the other hand, the existence of the neutral component
with a similar column density is controversial (e.g. Marthur 1997).
 
Conway \& Blanco (1995) used the VLA to obtain the H\,{\sc i} column 
density $N_{H\,{\sc i}}\sim 10^{22-23}\ {\rm cm^{-2}}$ toward the 
compact nucleus of Cyg~A, by measuring the 21~cm absorption. The neutral 
hydrogen is concentrated near the nucleus within several parsecs and
they suggested that atomic hydrogen is an important constituent of
the molecular torus obscuring the BELR of the narrow line radio galaxy.
Their result is compatible with an X-ray spectroscopy by the Ginga
satellite (e.g. Ueno et al. 1994) suggesting a highly optically thick component
around the compact nucleus.
 
Strongly polarized broad emission features around
$\lambda\lambda 6830,\ 7088$ in a fair fraction of symbiotic stars
were recently identified to be Raman-scattered O\,{\sc vi} 
$\lambda\lambda 1032,\
1038$ by atomic hydrogen of large column density (e.g. Schmid 1989,
Nussbaumer et al. 1989). The absorbed UV photons with wavelength $\lambda_i<
\lambda_\alpha=1216~{\rm\AA }$ by a hydrogen atom in the ground state $1s$
can be re-emitted either coherently with wavelength $\lambda_f =\lambda_i$
(the Rayleigh scattering) or incoherently with $\lambda_f=
(\lambda_i^{-1}-\lambda_\alpha^{-1})^{-1}$ (the Raman scattering).
Therefore, UV continuum photons near Ly$\beta$ are scattered and
transformed into optical photons around H$\alpha$ by the Raman
process. One of the most important characteristics of the Raman scattering
process is the enhancement of the Doppler shift so that the scattered
waves appear to be shifted by a factor of $\sim \lambda_f/\lambda_i$, and
this leads to a formation of a very broad line feature from a moderately
broad incident radiation source.
 
The Raman process can
be important in the systems having both a strong UV incident radiation source
and a neutral hydrogen component of large column density.
The optimum conditions for the Raman scattering may be provided by
AGNs, where the spectral energy distribution (SED)
is characterized by the big blue bump in the UV range and neutral components
with a high column density may be found near the broad line region. As in
symbiotic stars, the Raman scattering may produce observable features
around H$\alpha$ both in the profile and in the polarization.
 
In this paper we investigate the possible observational features
expected if the Raman process operate in AGN.
The paper is composed as follows. In section 2, we give a model description
and make an estimate of the strength of the Raman-scattered flux
using a standard AGN model. In section 3, the Monte Carlo result is
presented and compared with the electron-scattering model. We discuss
the application of the Monte Carlo result to a couple of narrow line AGN
and the observational implications in section 4. We summarize in the final
section.

\section{Model}

\subsection{Scattering by a high column neutral hydrogen}

The polarized fluxes of many Seyfert 2 galaxies are characterized by a constant
position angle and a constant degree of polarization, which strongly indicate
that they have the electron scattering origin. The width of the broad polarized
flux puts a constraint on the electron temperature $T_e \la 10^5\ {\rm K}$.
Assuming a covering factor $\ga 0.1$ and a similar luminosity of the hidden
broad line region of Seyfert 2 galaxies to that of Seyfert 1 galaxies,
one can deduce that
the electron column density is of order $10^{22}\ {\rm cm^{-2}}$ (e.g. 
Miller \& Goodrich 90).
 
Recent X-ray observations on several
Seyfert galaxies have shown that there exists a highly ionized component
of column density $10^{21}-10^{22}\ {\rm cm^{-2}}$. The warm absorber model
has been introduced to account for the X-ray absorption as well as
the UV absorption properties of the Seyfert galaxies (e.g. Marthur 1997,
Kriss et al. 1995).
Therefore the warm absorber may provide a good candidate for the
free electron scatterers of the UV photons originating from the broad line
region.
 
A high column neutral component has been suggested from
X-ray observations of Seyfert galaxies (Turner et al. 1996, Nandra et al. 1994,
Netzer 1993). The physical nature of the X-ray absorbing material in AGN
is still controversial, and the possibility that both
cold and warm absorbers may co-exist is not excluded.
 
Many X-ray observations of Seyfert 2 galaxies show that Compton reflected
components seen in the spectra in the band 0.1 keV - 10 keV are
compatible with the assumption of the molecular torus supplying the
cold optically thick material. It is not certain whether the neutral
absorbers suggested for the X-ray observations of Seyfert 1 galaxies
are also originated from the molecular torus or constitute an
independent component in the center part of AGN.
 
More direct information about the amount of neutral hydrogen can be
obtained from radio observations using the 21 cm line from the hyperfine
structure of hydrogen atom. Dickey (1986) investigated the content of neutral
hydrogen in 19 active spiral galaxies and obtained values in the range
$N_{H\,{\sc i}}\sim 10^{20-22} (T_s/100\ {\rm K})\ {\rm cm^{-2}}$.
He also mentioned the possibility that the spin temperature $T_s$ may
reside in a large range $10^{2-4}\ {\rm K}$, in which case a column density
as high as $10^{23}\ {\rm cm^{-2}}$ is also allowed.
 
Similar observations have been made by a number of researchers including
Gallimore et al. (1994), Mundell et al. (1995).
It seems that the results of these observations produce more or less
similar values for the neutral column density around the galactic nuclei.

In this subsection, we
investigate the scattering properties of atomic hydrogen. We will
assume a neutral component along the normal direction to the accretion
disk with a covering factor of about 0.1 and column densities $10^{21-23}
\ {\rm cm^{-2}}$.
 
Interactions of electromagnetic waves with atoms are described by
the Kramers-Heisenberg formula (e.g. Sakurai 1967).
The scattering types are divided into the Rayleigh scattering and
the Raman scattering, where in the former case the initial and
the final states of the scatterer are the same and the scattering is
coherent. On the other hand, in the case of the Raman scattering
the final state of the scatterer differs from the initial state and
therefore the scattered photon has different frequency from that of
the incident one. The explicit numerical computation of the relevant scattering
cross sections has been done by many researchers (Gavrila 1967,
Saslow and Mills 1969, Schmid 1989, Isliker et al. 1989, Sadeghpour \& Dalgarno
1992, Lee \& Lee 1997a).
 
Following Lee \& Lee (1997a), the Rayleigh and the Raman scattering cross
sections are given by
$$\eqalign{
{d\sigma_v\over d\Omega} &=r_0^2\ \left( {\omega'\over \omega}\right) \cr
&\times \Biggl|({\bepsilon}\cdot {\bepsilon'})
\Biggl(\sum_n  M^b_v(n) + \int_0^\infty dn' M^c_v(n') \Biggr) \Biggr|^2,
\cr} \eqno(2.1)
$$
where the subscript $v=\{Ray,\ Ram\}$ represents the scattering type,
$r_0={e^2\over m_e c^2}$ is the classical electron radius,
$\omega$ and ${\bepsilon}$ are the angular frequency and the polarization
vector for the incident wave and the primed quantities correspond to
the scattered wave.
The matrix elements for the bound states are
$$\eqalign{
M^b_{Ray}(n) &={1\over3}{m_e\omega\omega'\over \hbar}
<1s\parallel r\parallel np><np \parallel r \parallel 1s> \cr
&\times \left({1\over \omega_{n1}-\omega}+{1\over \omega_{n1}+\omega'}\right),
\cr} \eqno(2.2)$$
and
$$\eqalign{
M^b_{Ram}(n) &={1\over3}{m_e\omega\omega'\over \hbar}
<2s\parallel r\parallel np><np \parallel r \parallel 1s> \cr
&\times \left({1\over \omega_{n1}-\omega}+{1\over \omega_{n1}+\omega'}\right),
\cr} \eqno(2.3)$$
where the double-bar (or reduced) matrix elements for the position operator
are found in Berestetskii et al. (1972) (see also Bethe 1967).
Here, $\omega_{ni}
\{i=1,2\}$ is the transition angular frequency between the level $n$
and $1s$ or $2s$ depending on $i$.
The contributions to the matrix elements $M_{Ray}^c,\ M_{Ram}^c$ from
the continuum states are similarly given (e.g. Saslow \& Mills 1969).

\beginfigure*{1}
\vskip 91mm
\caption{{\bf Figure 1.}} Total (Rayleigh+Raman) and Raman scattering optical
depths for a neutral
hydrogen slab of column density $N_{H\,{\sc i}}=10^{22}\ {\rm cm^{-2}}$. The horizontal
axis shows the incident photon wavelength in units of \AA. The solid line shows
the total scattering optical depth and the Raman scattering optical
depth is represented by the dotted line. The dashed line represent the
ratio of the Raman scattering cross section to that of the total
scattering cross section. The points marked by a cross are the velocity
shifts having the total scattering optical depth of unity
for $N_{H\,{\sc i}}=10^{22}\ {\rm cm^{-2}}$, and the points marked by 
a circle are for $N_{H\,{\sc i}}=10^{23}\ {\rm cm^{-2}}$.
\endfigure

In Fig.~1, we show the total (Rayleigh+Raman) and the Raman scattering
optical depths
for a neutral hydrogen slab of column density $N_{H\,{\sc i}}=10^{22}\ 
{\rm cm^{-2}}$
as a function of the incident wavelength. We present the incident wavelength
$\lambda_i$ by the velocity deviation defined by
$\Delta V=c(\lambda_i-\lambda_\beta)/\lambda_\beta$.
A close inspection reveals that the total scattering cross section
is also slightly asymmetric. More quantitatively,
in the range $-850\ {\rm km~s^{-1}}< \Delta V < 1200\ {\rm km~s^{-1}}$,
the total scattering optical depth exceeds unity.
 
The ratio of the Raman scattering cross section to that of the
total scattering cross section is also plotted in Fig.~1.
It is particularly notable that the ratio is asymmetrical.
The ratio increases from $0.03$ to $0.3$ in the wavelength
interval $1010 ~{\rm\AA\ } <\lambda <1040~{\rm\AA}$.
This implies that the conversion efficiency from UV around Ly$\beta$ to optical
around H$\alpha$ is much larger in the red part than in the blue part.
 
From Fig.~1, we see that for a slab of column
$N_{H\,{\sc i}}=10^{23}\ {\rm cm^{-2}}$ incident photons with the 
velocity difference satisfying $-3300\ {\rm km~s^{-1}}<\Delta V<3100\ 
{\rm km~s^{-1}}
$ (or $1014~{\rm\AA }<\lambda< 1036~{\rm\AA }$) have total scattering
optical depth $\tau_{tot}>1$. The wavelength range considered in
this work is described by the Lorentzian wing part, and
the Doppler shift corresponding to the total scattering optical
depth of unity depends approximately on the square root of the column
density, i.e.,
$$|\Delta V | \sim 2000~N_{22}^{1/2}\ {\rm km~s^{-1}}, \eqno(2.4) $$
where $N_{22}=N_{H\,{\sc i}}/(10^{22}\ {\rm cm^{-2}})$.
 
The incoherent nature of the Raman scattering process introduces an
enhancement of the Doppler shift by the relation
$${\Delta\lambda_f\over\lambda_f} =\left({\lambda_f\over\lambda_i}\right)
\cdot \left({\Delta\lambda_i\over\lambda_i}\right), \eqno(2.5)$$
(e.g. Schmid 1989).
In the case of incident photons having wavelengths similar to that of Ly$\beta$,
the wavelength ratio $\lambda_f/\lambda_i \sim 6.4$, and therefore the incident
photons having a width $\sim 2000\ {\rm km~s^{-1}}$ around Ly$\beta$ will
result in a broad feature around H$\alpha$ with a width $\sim 13,000\ 
{\rm km~s^{-1}}$.
This implies that a broad line-like scattered feature with
$\Delta V\ga 15,000\ {\rm km~s^{-1}}$ can be formed from
a neutral hydrogen component with $N_{H\,{\sc i}} \ga 10^{22}\ 
{\rm cm^{-2}}$.
 
In this work, we do not consider the contribution of the H$\alpha$ line photons
scattered by hydrogen atoms in excited states $n=2$, because the velocity
width for this process to be important is a few times
the thermal velocity width depending on the population of the excited state
hydrogen atoms and is much smaller than the typical velocity scale of the
BELR. Therefore, this process has a very small band width compared with
that of the Raman scattering process. However, if we have a
neutral scattering component covering a large velocity space ranging $\ga
1000\ {\rm km~s^{-1}}$, then this process should not be neglected.
 
\subsection{Spectral Energy Distribution}

In order to assess the flux Raman-scattered by neutral hydrogen and compare
it with the flux scattered by free electrons, it is essential to
know the SED of the AGN both in UV and in optical ranges.
A typical AGN spectrum is characterized by the so called
big blue bump in the ultraviolet range and the specific luminosity
shows a peak at $\sim 10^{15}\ {\rm Hz}$ (e.g. Mathews \& Ferland 1987). It is
conventional to represent the SED as a power law with the power index
$\alpha$ so that we have
$$L_{\nu}=L_{\nu_0} \left({\nu\over \nu_0}\right)^\alpha. \eqno(2.6)$$
 
Zheng et al. (1997) investigated more than 200 quasar spectra
obtained from the Hubble Space Telescope (HST) and constructed composite
quasar spectra. They concluded that the power law index $\alpha=-0.99\pm0.05$
in the continuum between $1050~{\rm\AA}\ $ and $2200~{\rm\AA}$.
 
Natali et al. (1998) also investigated the optical-UV
continuum SED using the spectra of 62 QSOs emphasizing that the local
nature of the power-law approximation. They obtained the mean values of
the power-law index change from $\alpha\sim 0.15$ at $\lambda >3000~{\rm\AA}$
to $\alpha\sim -0.65$ at $\lambda <3000~{\rm\AA}$.
 
We assume that the incident spectrum in the UV and the optical ranges is
given by a double power law, i.e.,
$$L_\nu=\cases{L_{\nu_0} \left({\nu\over \nu_0}\right)^{\alpha_1}
\quad {\rm if} \quad \nu>\nu_0\cr
L_{\nu_0} \left({\nu\over \nu_0}\right)^{\alpha_2}
\quad {\rm if} \quad \nu<\nu_0,\cr} \eqno(2.7)$$
where $\nu_0=10^{15}\ {\rm Hz}$, and we set $\alpha_1=-1$,
and $\alpha_2=0.15$.
 
According to the prescription Eq.~(2.7), the incident photon number
flux $\Phi_\lambda$ is related with the specific luminosity
$$\Phi_\lambda\propto \lambda L_\lambda \propto\lambda^{-\alpha-1},\eqno(2.8)$$
and therefore, the continuum number flux near H$\alpha$ is given by
$$\Phi_{\lambda}(\lambda\sim\lambda_{H\alpha})\sim
0.4\times\Phi_{\lambda}(\lambda\sim\lambda_{L\beta})
\eqno(2.9) $$
 
Lee \& Lee (1997b) discussed the basic properties of the
Raman scattering process, and gave an empirical formula of the
efficiency $\epsilon_{Ram}$ for the conversion of the UV photons
into the optical photons.
They showed that the efficiency is quite large when the total
scattering optical depth is moderately large, or, $\tau_{scat}\ga 1$.
Furthermore, the conversion efficiency is sensitively dependent
on the ratio of the Raman scattering cross section to the total
scattering cross section.
 
Using the relation
$$d\lambda_f =\left({\lambda_f\over\lambda_i}\right)^2 d\lambda_i.\eqno(2.10)$$
and combining it with Eq.~(2.9), we obtain the Raman-scattered continuum
flux at around H$\alpha$
$$\eqalign{
\Phi^{Ram}_\lambda d\lambda_f
&\sim C_{N}\epsilon_{Ram} \Phi_{\lambda_i} d\lambda_i \cr
&\sim 0.06\ C_{N}\epsilon_{Ram} \Phi_{\lambda_f} d\lambda_f. \cr
}\eqno(2.11) $$
Here, $C_N$ is the covering factor of the neutral component having the
total scattering optical depth $\ga 1$, and $\lambda_f\sim 6563\ {\rm\AA\ }$,
$\lambda_i\sim 1025\ {\rm\AA\ }$, respectively.
 
The continuum number flux $\Phi_{Th}$ that is Thomson scattered near H$\alpha$
depends on the electron column density and the
covering factor. Hence, assuming that the scatterers have a small
Thomson scattering optical depth $\tau_W< 1$,
we may write
$$\Phi^{Th}_\lambda\sim C_{W}\tau_{W} \Phi_{\lambda_f},\eqno(2.12) $$
where $C_W$ is the covering factor of the ionized scattering medium.
Here, we ignore the line broadening by the thermal motions of the electrons.
 
Therefore, the Raman-scattered continuum flux around H$\alpha$ is compared
with the Thomson-scattered continuum flux by the relation
$$\Phi^{Ram}_\lambda \sim \epsilon_{Ram}\left({\tau_{W}\over 0.06}\right)^{-1}
\left({C_N\over C_W}\right) \Phi^{Th}_\lambda.  \eqno(2.13)$$
 
Under the assumption that the electron plasma has a column density
$10^{22}\ {\rm cm^{-2}}$ (or $\tau_W= 0.6\times 10^{-2}$), and that the neutral
hydrogen has the column of $N_{H\,{\sc i}}=10^{22}\ {\rm cm^{-2}}$, 
the continuum flux 
that is Raman-scattered from Ly$\beta$ with a width of $2000\ {\rm km~s^{-1}}$ 
is comparable to the Thomson-scattered at around H$\alpha$ with a width of
$15,000\ {\rm km~s^{-1}}$. Because $\epsilon_{Ram}$ is a sensitive
function of the column density and the frequency, the Raman-scattering
material can be regarded as a scattering-mirror with a narrow band width
centered at the resonance frequencies (e.g. Lee \& Lee 1997 a,b).
 
In an AGN spectrum, both Ly$\beta$ and H$\alpha$ are very strong lines and
especially Ly$\beta$ is often blended with O\,{\sc vi} 1034 to form 
a very broad
feature around 1030~\AA\ and thus the line photons are also an important
constituent of the UV incident radiation (e.g. Laor et al. 1994, Laor et al.
1995, Netzer 1990).
 
\subsection{Model and the Monte Carlo Method}
 
We briefly describe the model and the Monte Carlo code to compute
the polarization due to the neutral component. In this paper
the neutral scatterers are placed along the symmetry axis for
simplicity and ease of a direct comparison with electron-scattered
flux.
A more realistic model may be suggested by placing the neutral scatterers
in the molecular torus, the top part of which is accessible by an observer
with low latitude.
If the molecular torus is approximated by a perfect cylinder
with an inner radius $a$ and a height $h$ (e.g. Pier
\& Krolik 1992), then the total solid angle subtended by the accessible
part of the torus by an observer with the line of sight ${\bf\hat k_o}$
with respect to the light source at the origin is given by
$$\eqalign{
\Delta\Omega &=
{\mkern+40mu\raise.3ex\hbox{$\int\int$}\mkern-70mu\lower1.9ex\hbox{
$\scriptstyle{z\ge{{h\over2}+2a\cot\theta_0\cos\phi}}$}}
\ dz\ d\phi\ {a^2\over [a^2+z^2]^{3/2}} \cr
&=4\ \cot\theta_0\int_0^{\pi/2} du \ u \sin u \cr
&\times \left[1+\left({h\over 2a}
- 2\cot\theta_0\cos u\right)^2\right]^{-3/2}. \cr}
\eqno(2.14)$$
Here, we take the cylinder axis as $z-$axis and $\theta_0\equiv
\cos^{-1}{\bf\hat k_o} \cdot {\bf\hat z}$ is the
angle between the line of sight and $z-$axis. In Fig.~2a, a schematic
geometry is shown and the hatched region marked by `A' is the
accessible part by the observer.

\beginfigure*{2}
\vskip 91mm
\caption{{\bf Figure 2a.}} A schematic diagram for a cylindrical molecular torus
(see Pier \& Krolik 1992).
The inner radius is $a$ and the height is $h$. $z-$axis is chosen to
coincide with the symmetry axis and the central continuum source is
located at the origin. The hatched region marked
by `A' can be accessible by the observer with the line of sight ${\bf\hat
k_0}$. Assuming the the observer is in the $x-z$ plane, the region `A' is
described by the relation $z\ge {h\over 2a}+2a\cot\theta_0\cos\phi$, where
$\theta_0=\cos^{-1}{\bf\hat k_0} \cdot {\bf\hat z}$.
\endfigure

For parameters $h/2a = 1$ and $\theta_0=\cos^{-1}0.5$, a simple numerical
integration of Eq.~(2.14) gives
the solid angle $\Delta\Omega /4\pi = 0.08$. Hence, this particular
geometry mimics a neutral cloud located at high latitude to the observer
with the covering fraction $\sim 10$ percent.  However, a tapered disk
model investigated by Efstathiou et al. (1995) would give much smaller
solid angle. Therefore, only after a detailed geometry of the cicumnuclear
region is specified with the appropriate photoionization computation,
the covering fraction of the neutral scatterers may be assessed.
In this paper we choose a very simple geometry as depicted in Fig.~2b to find
the main features of Raman scattering by neutral hydrogen without invoking
many free parameters necessary to specify a detailed geometry.
In this geometry, the position angle of the polarized radition
is perpendicular to the symmetry axis, which is regarded as the radio
axis of the active nucleus.

\beginfigure*{3}
\vskip 91mm
\caption{{\bf Figure 2b.}} A schematic diagram for the scattering model.
The continuum light source lies in the center and the broad line photons
are generated in the vicinity of the continuum source. The neutral hydrogen
component is located along the $z-$axis, which is coincident with the normal
direction to the accretion disk.
\endfigure

As shown in Fig.~2b, the light source is
located at the center and blocked by an opaque component lying equatorially,
and the neutral hydrogen component lies alogn $z-$axis that is
coincident with the normal direction of the accretion disk.
The geometry of the scatterer is assumed to be an oblate spheroid with the
minor axis aligned to the $z-$axis. Along the major axes the column density
is assumed to be $2 N_{H\,{\sc i}}$ and inside the scattering region, 
the number
density is assumed to be fixed, which enables us to measure the physical
distance in terms of column density. The covering factor $C_N$ of the
neutral scatterers is set to be $0.15$.
 
We prepare the UV incident continuum in accordance with the double power
law given by Eq.~(2.7). We also prepare the line photons around
1030 \AA\  corresponding to the blended Ly$\beta$ and O\,{\sc vi}. 
We combine the three
gaussians centered at $1025\ {\rm\AA} $, $1032\ {\rm\AA} $ and
$1038\ {\rm\AA }$ with the same width of $3000\ {\rm km~s^{-1}}$ and strength
ratio of $1:1.5:1.5$. The total blended line flux is renormalized so that
it has an equivalent width of $30~{\rm\AA} $ (e.g. Korista et al. 1997).
The possibility for intrinsic anisotropy in the radiation
field in the BELR has been suggested and observational evidence is
presented for NGC 1068 (Miller et al. 1991, Hough and Young 1995). However
for simplicity it is assumed that the light source is isotropic.
 
As a UV photon reaches the scattering region, it may be first
Rayleigh-scattered several times before it is Raman-scattered. It is
assumed that the continuum opacity near H$\alpha$ is negligible and therefore
any Raman-scattered photon leaves the scattering region. Here, the thermal
motion of the scatterers is not important because the bulk velocity
of the BELR is much larger.
 
The polarization of the line photon resonantly scattered by a hydrogen
atom is dependent on the incident wavelength in the rest frame of the
scatterer. However, most UV photons are scattered in the damping wing
regime where the deviation from the resonance transition frequency is
sufficiently large to ensure that the scattering is characterized by the
classical Rayleigh phase function. This is a direct consequence that the
fine structure splittings in the $n p$ states are negligible compared with
the frequency deviation corresponding to the damping wing regime
(e.g. Stenflo 1980). Because the final states for both the
Rayleigh scattering and the Raman scattering are $s$ states, the same
Rayleigh phase function is used to obtain the wave vector and the
polarization associated with the scattered photon for both the Rayleigh
and the Raman scatterings.
 
\section{Results}
 
\subsection{Continuum and Basic Properties}

\beginfigure*{4}
\vskip 91mm
\caption{{\bf Figure 3.}} The scattered flux, the polarized flux and the linear 
degree of polarization around H$\alpha$ Raman-scattered by
a neutral hydrogen scattering component of column densities
$N_{H\,{\sc i}}=10^{21},\ 10^{21.5},\ 10^{22},\ 10^{22.5}\ 
{\rm cm^{-2}}$.
The UV incident spectrum is the continuum given by Eq.~(2.7),
the photons with wave vector ${\bf\hat k}$ satisfying
$0.45 < {\bf\hat k} \cdot {\bf\hat z} < 0.55$ are collected, where $z$ axis is
the symmetry axis of the AGN. The solid line represents the scattered flux
the dotted line stands for the polarized flux multiplied by 6 for comparison
with the scattered flux and the dotted line with error
bars for the degree of polarization.
The error bars are 1-$\sigma$ obtained from the Monte Carlo computation.
\endfigure

In Fig.~3, we show the scattered flux, the polarized flux and the linear
degree of polarization
around H$\alpha$ resulting from Raman scattering by neutral hydrogen of column
densities in the range $N_{H\,{\sc i}}=10^{21-22.5}\ {\rm cm^{-2}}$. The
UV incident spectrum consists of only the continuum given by Eq.~(2.7), and
we collect the photons with wave vector ${\bf\hat k}$ satisfying
$0.45 < {\bf\hat k} \cdot {\bf\hat z} < 0.55$, where $z-$axis is the symmetry
axis of the AGN.
 
The polarized flux appears to be a normal H$\alpha$ line with
$$FWHM \sim 1.6\times 10^4\ (N_{H\,{\sc i}}/10^{22}\ 
{\rm cm^{-2}})^{1/2}\ {\rm km~s^{-1}}. \eqno(3.1) $$
This corresponds to a few times the width shown in Eq.~(2.4).
The scattered flux shows a peak at the line center,
and has a red excess that becomes more conspicuous as the column density
increases. The polarized flux is broader than the scattered flux
and shows a red-shifted peaks. The locations of the peak of the polarized flux
are given by
$\Delta V_{peak}\sim 5600\ {\rm km~s^{-1}}$ for $N_{H\,{\sc i}}=
10^{22.5}\ {\rm cm^{-2}}$, and
$\Delta V_{peak}\sim 1700\ {\rm km~s^{-1}}$, and $\sim 800\ {\rm km~s^{-1}}$
for $N_{H\,{\sc i}}=10^{22}\ {\rm cm^{-2}}$ and $10^{21.5}\ 
{\rm km~s^{-1}},$ respectively.

Near the line center, the scattering optical depth is very large, and
therefore a large scattering number is needed to escape from the
scattering region. As the scattering number gets larger, the conversion
from Ly$\beta$ to H$\alpha$ becomes more efficient, and hence we obtain
a stronger Raman flux.
 
Furthermore, the conversion efficiency is dependent on the ratio of the
Raman scattering cross section to that of the total scattering cross section
in a highly sensitive way. Therefore, as the column density becomes large,
the mean scattering number also increases and the strength of the
Raman scattered flux is enhanced preferentially to the red part
where the ratio of the Raman scattering cross section is larger
as shown in Fig.~1.
 
An illustrative explanation is given by Lee \& Lee (1997 b) in their analysis
of the Raman scattered flux in symbiotic stars. It is notable that the red
asymmetric polarized profile in the high column case is originated from the
atomic physics rather than extrinsic factors such as kinematics.
 
On the other hand, the largest linear degree of polarization is usually
obtained when the mean scattering number is $\sim 1$, because a large number
of scatterings randomize the radiation field associated with the scattered
photons and no scattering yields any polarization. However, in the
case of the Raman scattering, where the scattered photons are not mixed with
the unscattered flux due to the incoherence, the degree of polarization
monotonically decreases as the mean scattering number increases. Therefore,
the polarization tends to decrease as the total scattering cross section
increases.
 
Fig.~3 shows that irrespective of the column density, the degree of
polarization is the smallest at the H$\alpha$ center, where the total scattering
cross section is the largest. It is also interesting that the minimum degree
of polarization at the center is approximately constant independent of the
scattering column density. This behavior is explained by noting that
the Raman flux is mainly contributed from the first few scatterings and that
negligibly polarized Raman photons contribute little because of low
probability of surviving a large number of Rayleigh scatterings. Near the
center, the total scattering optical depth is large enough even for 
$N_{H\,{\sc i}}=
10^{21}\ {\rm cm^{-2}}$ that the polarized flux and the degree of polarization
become saturated.
 
Because the total scattering optical depth rapidly decreases as the wavelength
deviation from the center wavelength gets larger, the scattered flux becomes
negligible. This is also illustrated by the large error bars in the figure.
It is interesting  to note that the polarized flux, the total
scattered flux and the linear degree of polarization are dependent
on the wavelength and the column density of the scatterer, as
may be manifested by fine structures in the spectropolarimetric observations.
 
\subsection{Line Contribution and Electron Scattering}

\beginfigure*{5}
\vskip 91mm
\caption{{\bf Figure 4.}} The same quantities as in Fig.~3 except for the
UV incident radiation includes the line contribution from the Ly$\beta$ and
O\,{\sc vi} doublet.
\endfigure

In Fig.~4 are shown the same quantities as in Fig.~3 except for
the UV incident radiation that includes the line contribution from the 
Ly$\beta$ and O\,{\sc vi} doublet. A gaussian profile for each line
is assumed. With a continuum flux $F_c$ and a line flux
$$F_l = F_0 \exp[-(\lambda-\lambda_0)^2/2\sigma_l^2] \eqno(3.2)$$
the equivalent width $EW$ is given by
$EW=\sqrt{\pi}\sigma_l (F_0/F_C)$.
We take $\sigma_l=\lambda_0 (3000\ {\rm km~s^{-1}}/c)$ for each line.
The equivalent widths of the components of the blended feature are given by
$EW_{Ly\beta}= 7.5\ {\rm \AA } $, $EW_{O\,{\sc vi}}= 22.5\ {\rm \AA } $,
where the doublet O\,{\sc vi} is assumed to be contributed equally from
$\lambda 1032$ and $\lambda 1038$ components.
 
The qualitative features shown in Fig.~4 are similar to those in Fig.~3.
The main difference is the strength of the polarized flux, due
to the increase of the incident UV photons by a factor of $\sim 2$.
Because the Raman-scattering mirror has a narrow band
width, the overall variation of the incident UV flux over this mirror-width
is small and therefore we obtain qualitatively similar polarized fluxes.
 
However, because the blended line feature is centered near 1030 \AA,
the red side of Ly$\beta$ is stronger than the blue side by a small amount.
This leads to more red-asymmetric Raman-scattered flux, which is barely
apparent in the figure. The small effect of O\,{\sc vi} is due to 
the small scattering cross section, and the major contribution to the polarized
flux is provided by the Ly$\beta$ line photons which possess a large scattering
cross section. When $N_{H\,{\sc i}}\la 10^{21}\ {\rm cm^{-2}}$, 
the Raman scattering is
not important and the asymmetry in the profile is negligible.

\beginfigure{1}
\vskip 91mm
\caption{{\bf Figure 5.}} The electron-scattered flux, the Raman-scattered flux
and their sum are represented by the dotted line, the dotted line
with crosses and the solid line. The Raman-scattered flux is the
same as in Fig.~4 and shown for comparison. The long dashed line
stands for the synthetic linear polarization of the total
scattered flux diluted by an unpolarized continuum
$\Phi_{uc}=5\times 10^{-3} \Phi_\lambda$, where
$\Phi_\lambda$ is given by Eq.~(2.4).
\endfigure

In Fig.~5 we compare the Raman-scattered polarized flux with the polarized flux
scattered by an electron plasma located in the same position with the
same covering factor as the neutral hydrogen scatterer. The column density
$N_e$ of the free electron is assumed to be $N_e = 10^{22}\ {\rm cm^{-2}}$.
The incident flux consists of the continuum
described in Eq.~(2.7) and the broad H$\alpha$ given by a gaussian with
$\sigma_l=3000\ {\rm km~s^{-1}}$ and the equivalent width $EW_{H_\alpha}
=100\ {\rm \AA}$.
 
If we assume that the thermal velocity of the electron plasma is much
smaller than the width of the broad Ly$\beta$, then the line broadening
by thermal electrons is neglected.
If the Thomson scattering optical depth is also small, then the
Thomson scattered flux is mainly contributed from singly scattered
photons. Under these assumptions, the electron-scattered flux is characterized
by a constant degree of polarization and
therefore, the polarized flux has almost the same shape as the
that of the scattered flux
 
Therefore, in this optically thin limit, the degree of
polarization $p_{Th}$ of the Thomson scattered flux is given by
$$p_{Th} = {1\over 2\pi(1-\mu_e)}\int_{\mu_e}^{1}\int_0^{2\pi} d\mu\  d\phi
\  {p_1 - p_2 \over p_1 + p_2}           \eqno(3.3)$$
where $p_1 = (\cos\theta_o\sin\phi)^2+(\cos\theta_o\cos\theta\cos\phi
+\sin\theta_o\sin\theta)^2$ and $p_2=\cos^2\phi +\cos^2\theta\sin^2\phi$
are the scattered components that are polarized parallelly and perpendicularly
to the plane spanned by the observer's line of sight and the symmetry
axis ($z-$axis) of the system, respectively (e.g. Brown \& McLean 1977).
Here, $\theta_o$ is the angle between the $z$ axis and the observer's line
of sight, $\mu_e$ is the cosine of the half-angle of the conical region
which the electron plasma subtends with respect to the source.
 
For $\theta_o=\cos^{-1}0.5$ and $\mu_e = 0.7$, a simple numerical
integration gives $p_{Th}=0.39$, which agrees well with the
largest degree of polarization obtained for the Raman scattered
flux shown in Fig.~3.
 
The electron-scattered flux around H$\alpha$ is comparable with the Raman
scattered flux for $N_{H\,{\sc i}}=10^{21.5}\ {\rm cm^{-2}}$. 
We include an unpolarized
continuum of strength $\Phi_{uc}=5\times 10^{-3} \Phi_\lambda$
as suggested by Tran (1995 c) to compute the synthetic linear degree of
polarization represented by the long-dashed line in Fig.~5.
With the polarization-diluting continuum, the degree of polarization
shows a bump-like feature around H$\alpha$, observed in a number of Seyfert
2 galaxies (e.g. Tran 1995 a,b,c). It is particularly noted that the
polarization bumps also have a dip in the center which is ascribed to the
decrease of polarization arising from a large number of Rayleigh-Raman
scatterings.
 
\section{Observational Implications}
 
As mentioned in section~2.2, the Raman-scattering
material may be regarded as a scattering-mirror with a very narrow band width
centered at the resonance frequencies and furthermore the band width
provides a natural measure of the column density of the neutral hydrogen.
Therefore, when the incident flat continuum is reflected by the
Raman-scattering mirror, the scattered continuum appears a line-like
feature around the transition frequency $\nu_i-\nu_{Ly\alpha}$.
 
Stark et al. (1992) presented a H\,{\sc i} survey of our galaxy 
using a beam with
a size of FWHM$=2^\circ$ and showed that the H\,{\sc i} column 
density toward the
galactic center amounts to $10^{22}\ {\rm cm^{-2}}$. Gallimore et al. (1997)
presented the H\,{\sc i} column density toward the radio jet of Mrk 6,
and showed that neutral hydrogen of column $\sim 10^{21}\ {\rm cm^{-2}}$
is present around the active nucleus.
X-ray observations of many AGN reveal that X-ray absorbing material
of $N\sim 10^{20-23}\ {\rm cm^{-2}}$ around the nuclear region and the
physical nature, the exact location and kinematics are currently
being investigated intensively.
 
Ogle et al. (1997) performed a spectropolarimetry of the narrow line radio
galaxy Cyg A and found that the full width at half maximum of the polarized
H$\alpha$ flux is $26000\ {\rm km~s^{-1}}$. They noted that the
broad emission lines of some radio loud active galaxies show a large
velocity width up to $30000\ {\rm km~s^{-1}}$ with double-peaked profiles
(e.g. Eracleous \& Halpern 1994). However, Antonucci et al. (1994) observed
the same object and measured the width of the broad emission line
Mg~II  $\sim 7500\ {\rm km~s^{-1}}$, which is 3.5 times smaller than that
of the polarized H$\alpha$.
 
The thermal width of an electron plasma with the temperature $T_e$
is given by
$$v^e_{th}\sim 400\ T_{e4}^{1/2}\ {\rm km~s^{-1}},\eqno(4.1)$$
where $T_{e4}=T_e/(10^4\ {\rm K})$.
If the width of the H$\alpha$ in the polarized flux is
interpreted to be caused by hot electrons, then the temperature of
the electron plasma $T_e \sim 5\times 10^{7}\ {\rm K}$ (e.g.
Weymann 1970). And there may also exist other polarized lines with a similar
width, and it is difficult to explain the width of Mg~II line observed
by Antonucci et al. (1994).
 
On the other hand, if the Raman-scattering processes by the neutral
component operate, then it may provide an interesting interpretation,
i.e., the large width may not be associated with the Doppler effect but
mainly an effect of the atomic physics associated with the Raman scattering.
 
Conway \& Blanco (1995) measured the column density of H\,{\sc i} 
toward the compact
radio nucleus of Cyg A and obtained a value $\ge 2.54\pm 0.44\times
10^{19}\ T_s \ {\rm cm^{-2}}$, where the spin temperature $T_{s}$
may go up as high as $1000-8000\ {\rm K}$ depending on the physical state of
the putative molecular torus obscuring the compact nucleus. Therefore
the existence of a high column neutral hydrogen with
$N_{H\,{\sc i}}\ga 10^{22}\ {\rm cm^{-2}}$ is not ruled out. The 
Raman-scattering
by this component may introduce a broad polarized flux with width
$\sim 2\times 10^4\ {\rm km~s^{-1}}$ redshifted by $\sim 2000\ {\rm km~s^{-1}}$
from incident UV radiation around Ly$\beta$ with width $\ge 3000\ 
{\rm km~s^{-1}}$.
Therefore, if we take the width of Mg II 2800 as a representative velocity
scale of the BELR, then the spectropolarimetry of Cyg~A by Ogle et al. (1997)
is consistent with this interpretation.
 
Young et al. (1993) presented their spectropolarimetric observation of the
Seyfert 2 galaxy IRAS~110548-1131 (see also Young et al. 1994). They found
a very broad H$\alpha$ line in the polarized flux with a full width at zero 
intensity of $16800\ {\rm km~s^{-1}}$ and FWHM$=7600\ {\rm km~s^{-1}}$, 
which, they 
note, is wider than that of most Seyfert 1 galaxies. Another interesting point 
is that the peak of the broad polarized
H$\alpha$ is red-shifted by $900\pm 400\ {\rm km~s^{-1}}$ with respect to the 
narrow H$\alpha$ emission, which they ascribe to an outflow of the scatterers.
 
If we apply our result, the most plausible column density for
IRAS~110548-1131 is
$\sim 10^{21.5}\ {\rm cm^{-2}}$, which gives the FWHM of the polarized flux
$\sim 10^4\ {\rm km~s^{-1}}$ and the redshift of the peak $\sim 800
\ {\rm km~s^{-1}}$.
 
However, there are other factors that may possibly affect the redshift of the
polarized flux.
For example, the broad lines are usually redshifted with respect to the
narrow emission lines in many AGNs. The amount of redshift differs for
various lines, and this may be associated with the stratified structures
of the BELR consistent with the reverberation studies of several Seyfert
galaxies (e.g. Peterson 1993).
In addition to the velocity shift in the source part,
the kinematics of the scattering agents is important to determine
the overall profile of the scattered (and polarized) flux.
In particular, Conway and Blanco (1995) pointed out that the H\,{\sc i} 
component in
Cyg A is moving with a typical velocity of several hundred ${\rm km~s^{-1}}$.
We omit the quantitative analysis of this effect in this paper.
 
The modelling by Young et al. (1996) indicates that the broad lines
of IRAS~110548-1131 are polarized by the same degree as the continuum,
for which case the Raman scattering is not required. Therefore,
if the Raman scattering also operates, then the Raman scattering features
may be found only in the fine structures in the polarized flux with
high resolution and good signal to noise ratio requiring large amount
of integration time.
 
Here, one example of the fine structures in the
polarized flux can be a relative shift of the narrow line peak and that
of the polarized flux.
The electron-scattered flux reduces the asymmetry in the total
scattered flux if the broad H$\alpha$ is symmetric. Therefore
the relative strength of the electron (and/or dust)-scattered
flux affects the location of the polarized flux peak.
 
Another possible solution may be provided by observing metal lines
including Mg~II 2800 in the spectropolarimetry mode. Because metal lines
have negligible scattering cross section, they will not be scattered
by atomic hydrogen and may possess the polarized flux
entirely due to free electron. Therefore, the Raman scattered Balmer
lines can be shown by verifying different shapes and widths of
the polarized fluxes compared with those of metal lines.
 
One important process expected from a high column neutral hydrogen
is the Rayleigh scattering of the photons around the Lyman series lines. The
Rayleigh scattering of Ly$\alpha$ photons has been discussed in symbiotic
stars by Isliker et al. (1989). Because the Rayleigh
scattering cross section is in general larger than the Raman scattering cross
section, it is expected that the Lyman lines are almost as strongly polarized
as the Balmer lines.
 
In this respect, an interesting observation is the HST spectropolarimetry on
the quasar PG~1630+337 by Koratkar et al. (1995), who found out a strong
polarized flux around Ly$\alpha$. It is a theoretical possibility that
this feature is formed by the Rayleigh scattering process, in which case
a more weakly polarized Ly$\beta$ flux is expected. Because the strong broad 
line N~V 1240 is present at $\sim 6000\ {\rm km~s^{-1}}$ to the red side of 
Ly$\alpha$, the
possibility that the polarized flux is mainly contributed from receding N~V is
not excluded as noted by Koratkar et al. (1995) (see also Arav et al. 1995).
 
Therefore UV spectropolarimetry on nearby AGN and optical spectropolarimetry
of high redshift AGN will provide strong constraints on the importance
of neutral hydrogen scatterings.
 
\section{Summary}
 
We performed a Monte Carlo computation of the linear polarization
expected from the Raman scattering process by a neutral hydrogen component
in active galactic nuclei. The main features of the Raman
scattering processes in AGN include a large width due to the enhancement
of the Doppler shift, and a possible red asymmetric polarized flux around 
H$\alpha$ ascribed to the asymmetric conversion efficiency with a smaller 
contribution of Ly$\beta$ and O\,{\sc vi} blended broad emission lines.
 
It is not still clear whether the neutral component
with a high column density exists in a typical AGN, and remains a very
intersting possibility that the Raman scattering processes provide
a natural interpretation of several spectropolarimetric observations
of narrow line AGN including Cyg~A and IRAS~110548-1131.
It is concluded that more detailed comparisons of polarized flux of
other metal lines and hydrogen Balmer lines will put stronger constraints
on the possible effects of high column neutral components in AGN.
 
\section*{Acknowledgments}

HWL thanks Ki-Tae Kim for providing the information on
the H\,{\sc i} distribution of our galaxy and external galaxies.
He is also grateful to Sang-Hyeon Ahn for helpful discussions about
AGN models. JHY wants to express his gratitude to Chan Park for
kind advice on fortran programming.
We thank the anonymous referee for many useful and
important comments which improved this paper.


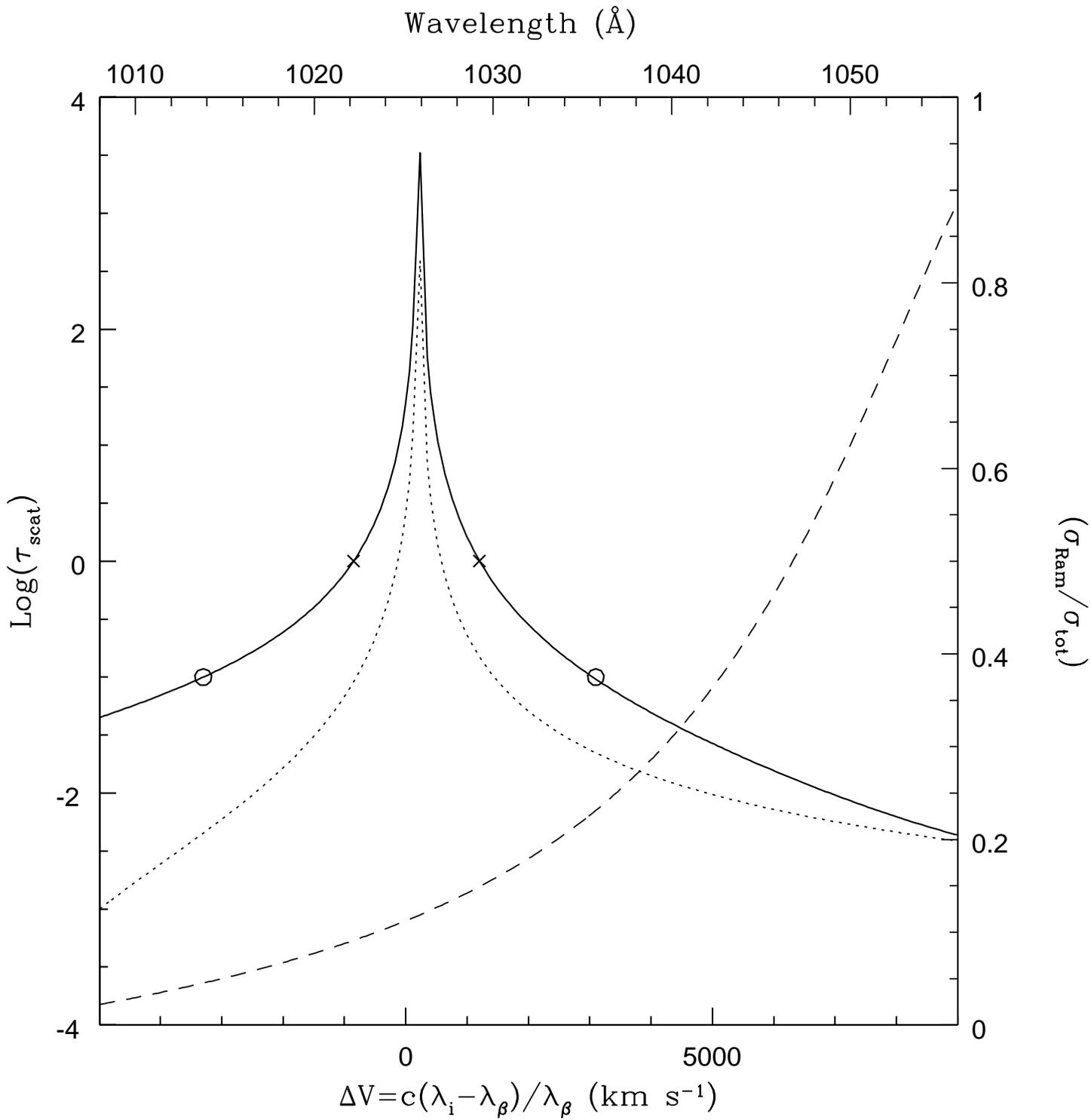
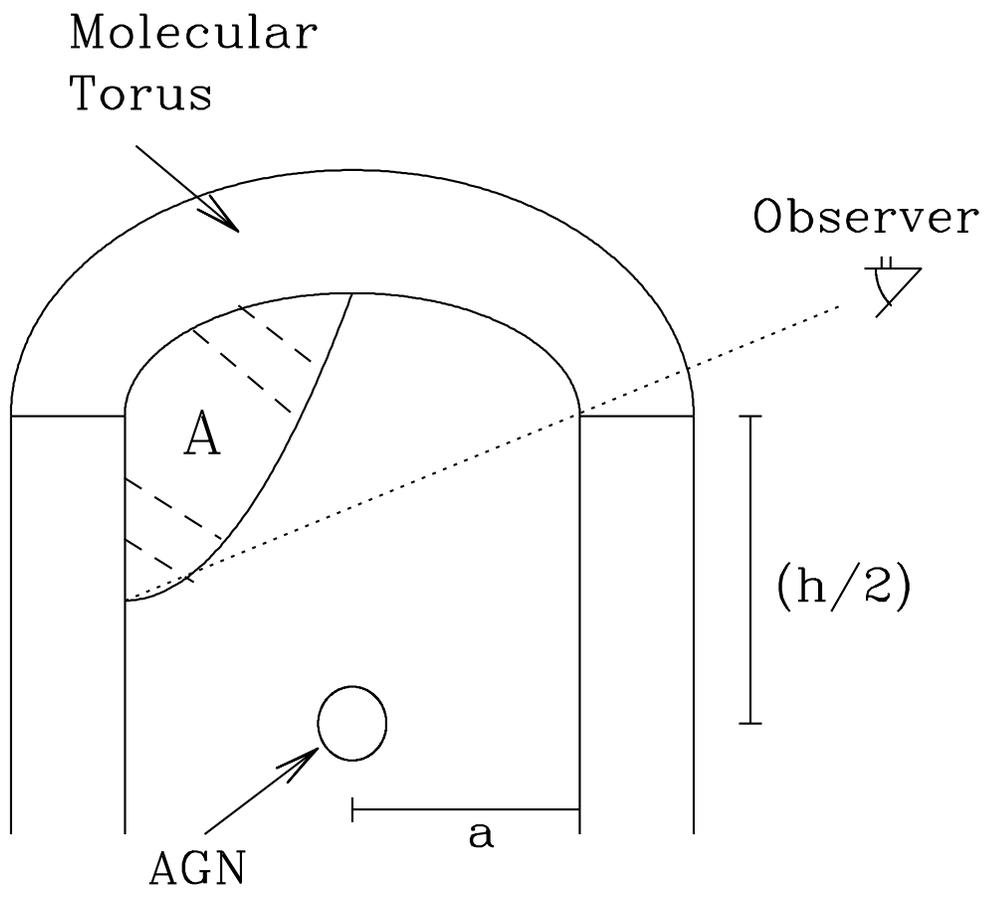
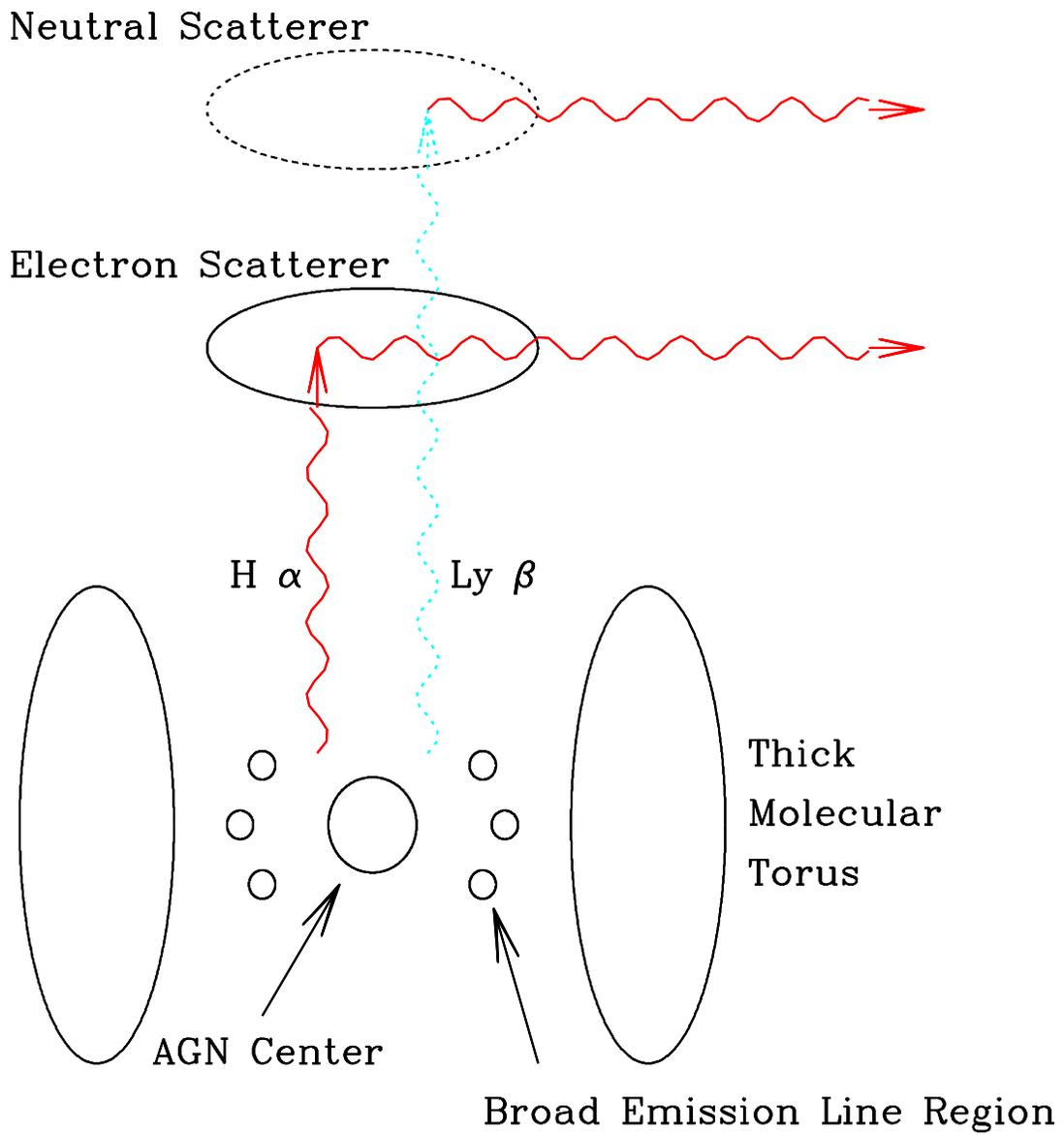
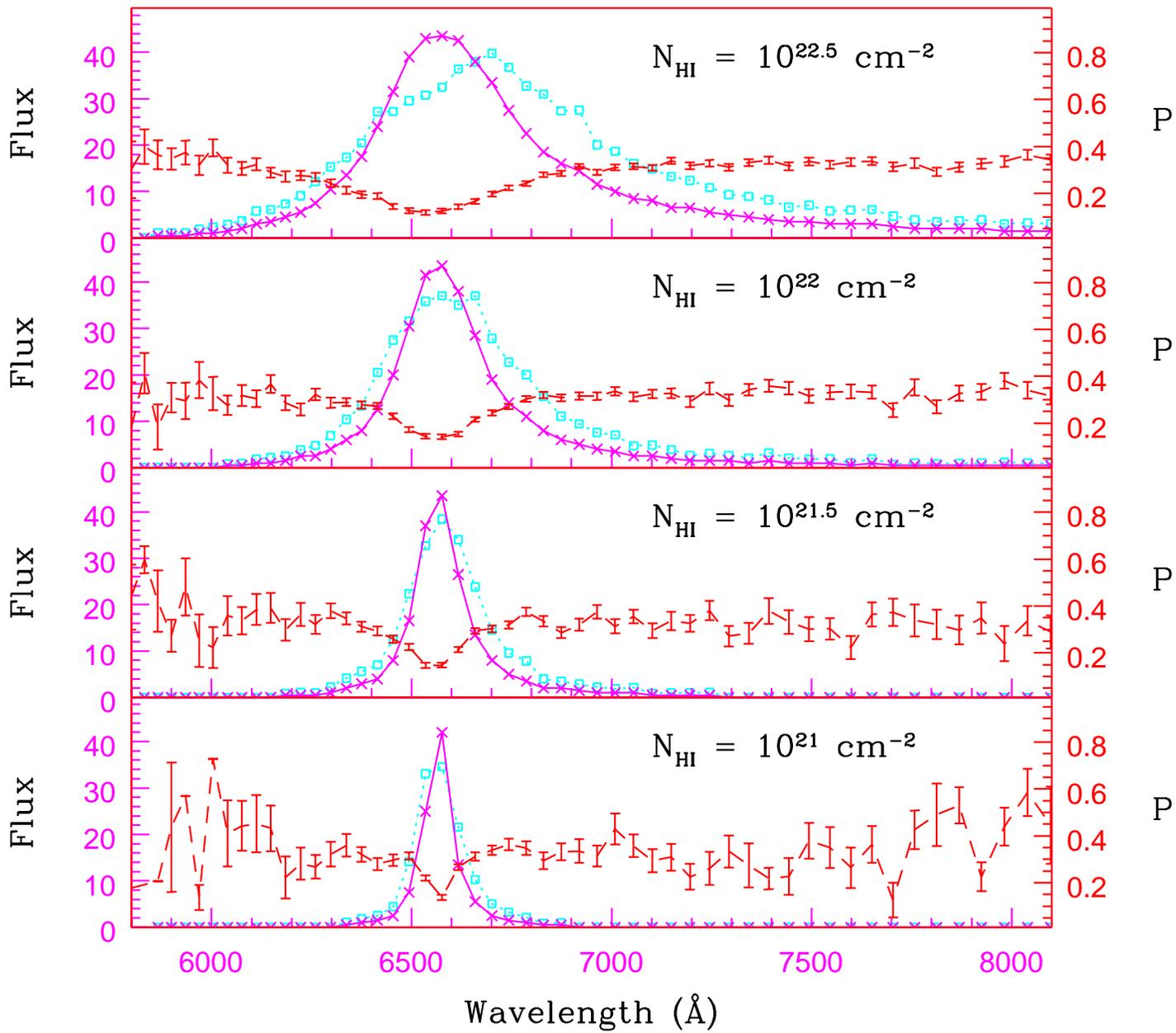
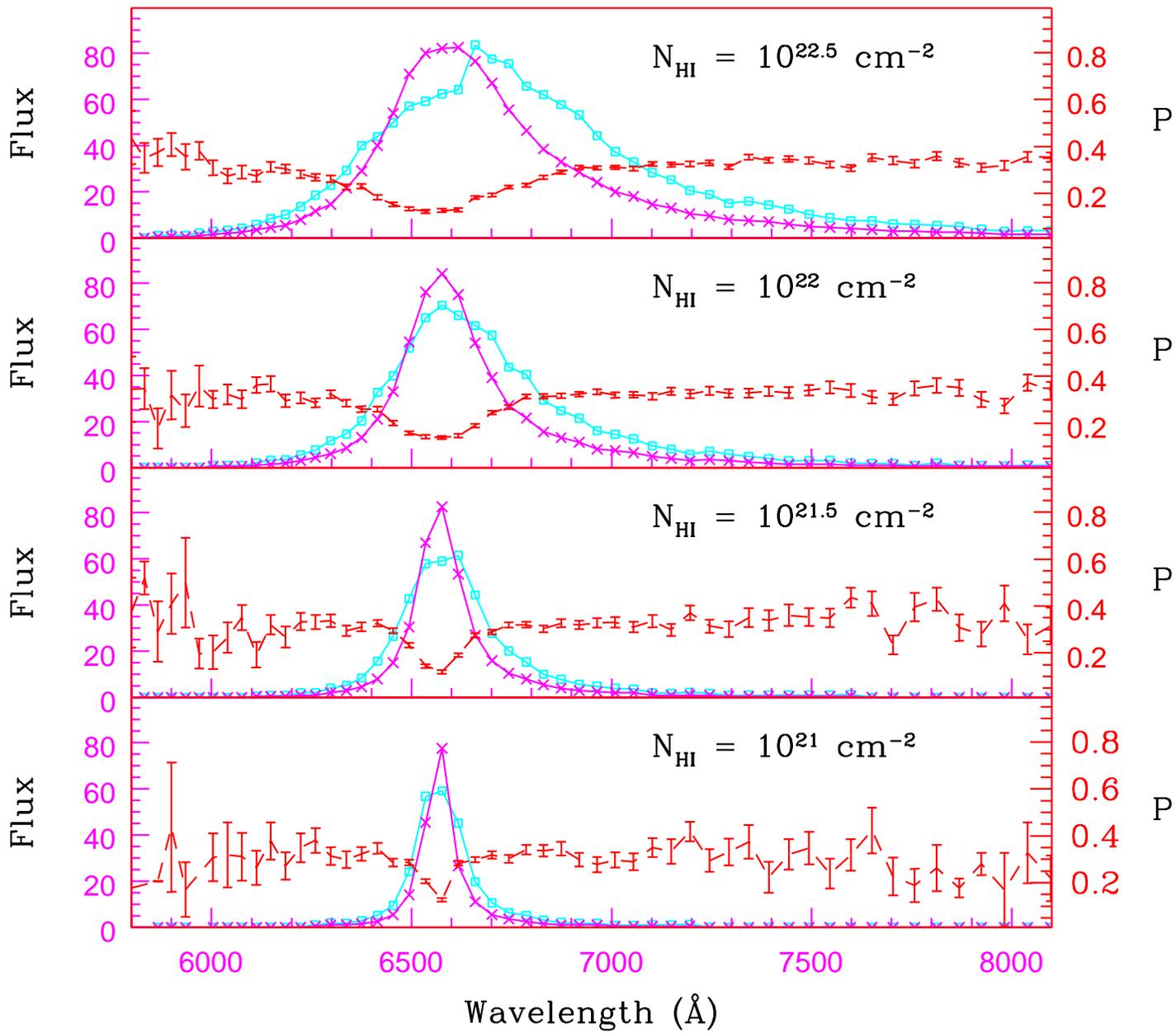
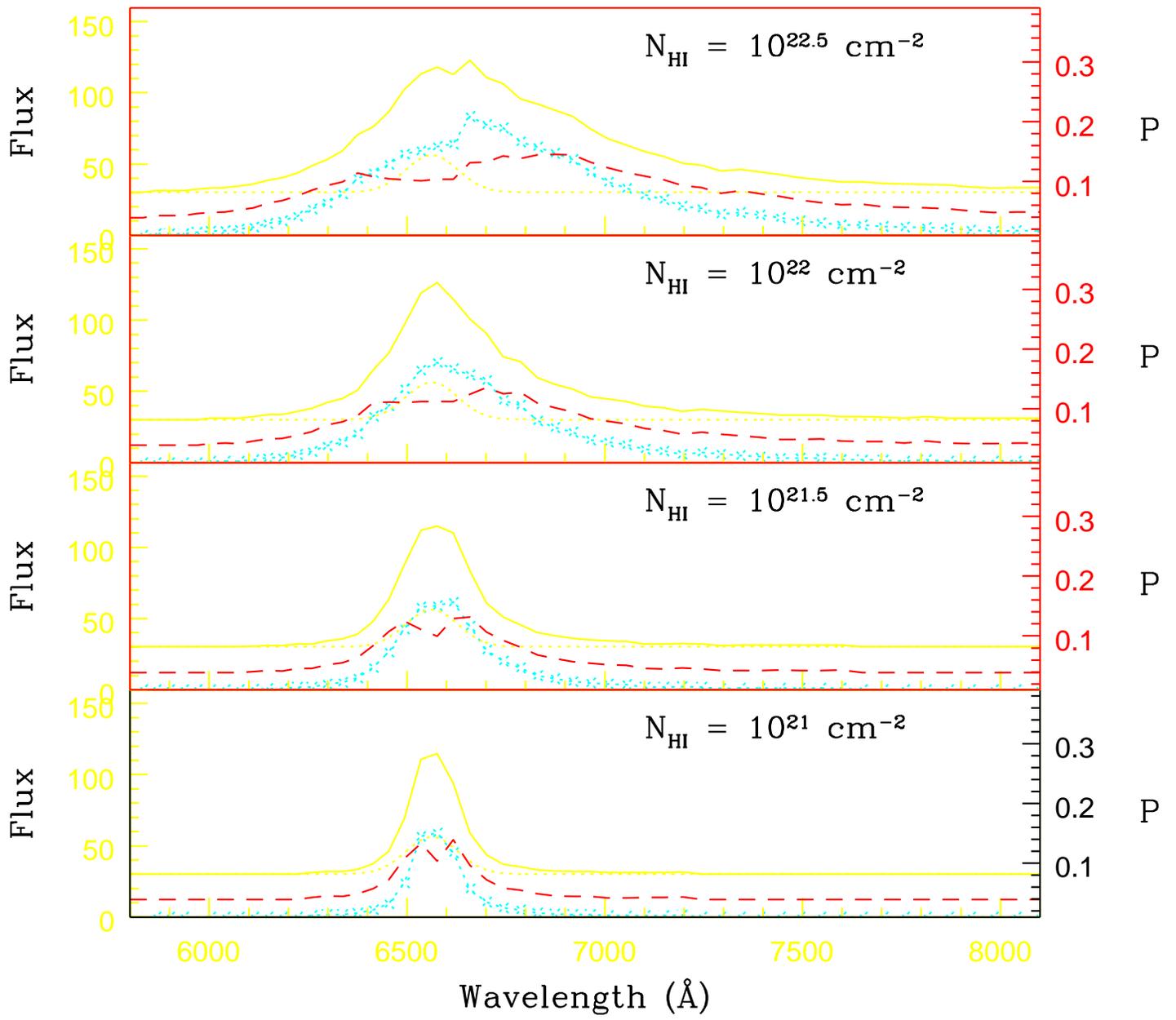
\bye